\documentclass[twocolumn,showpacs,preprintnumbers,amsmath,amssymb]{revtex4}

\usepackage{graphicx}% Include figure files
\usepackage{dcolumn}% Align table columns on decimal point
\usepackage{bm}% bold math
\usepackage{color}
\usepackage{amsmath}
\usepackage{amssymb}
\usepackage{dsfont}
\usepackage{color}
\usepackage{calc}
\usepackage{hyperref}
\usepackage{multirow}
\usepackage{subfigure}
\usepackage{float}
\usepackage[normalem]{ulem}
\usepackage{natmove}

\def\be{\begin{equation}}
\def\ee{\end{equation}}
\def\bea{\begin{eqnarray}}
\def\eea{\end{eqnarray}}

\graphicspath{{./figures/}}

\begin{document}

\title{Self similarity of the expanding universe as understood by quantum-phase-fields}

\author{~I.~Steinbach}
\email{ingo.steinbach@rub.de}
\author{~J.~Kundin}
\email{julia.kundin@rub.de}
\author{~F.~Varnik}
\email{fathollah.varnik@rub.de}
%\email{fathollah.varnik@rub.de}

\affiliation{ Ruhr-University Bochum, ICAMS, Universitaetsstrasse 150, DE-44801 Bochum, Germany}
%\author[icams]{Ingo Steinbach}


%\address[icams]{Ruhr-University Bochum, ICAMS, Universitaetsstrasse 150, Bochum}

\date{\today}% It is always \today, today,

\begin{abstract}
An idea of the universe as a self-contained system of interacting fields as a closed doublon network is developed. The characteristic scale of this system is considered emergent from general principles. Self similarity of patterns in the universe can be understood by introducing a characteristic scale with respect towhich patterns are evaluated. The developed physical-mathematical model is in  good agreement with cosmological length scales and gives a rationale of the empirical fact for an expanding universe. Within the framework of the de~Broglie-Bohm double solution program, the model can be applied to rationalize the existence of a ground potential and a cosmological constant.
\end{abstract}

%\pacs{03.70.+k, 04.50.Kd, 05.70.Ln}% PACS, the Physics and Astronomy

\maketitle
%{Corresponding to the published version in: Zeitschrift f\"ur Naturforschung A; 72, 51-58 (2017); DOI 10.1515/zna-2016-0270, with several formal corrections.}

\section{Introduction}\label{Intro}

Each categoric feature of our physical world is commonly attributed by a typical `scale'. In the anthropic environment we use units related to the human body, like meter and feet, seconds ($\approx$ one heart beat), pounds or kilograms corresponding to our body weight. In rational physics we use Planck units. They are defined exclusively in terms of five universal physical constants: Speed of light $c$, Planck's constant $\hbar $, vacuum permittivity $\epsilon_0$, Boltzmann constant $k_B$ and the coefficient of gravitation $G$.  In the context of this essay we will deal only with speed of light $c$, Planck's constant $\hbar$ and the coefficient of gravitation $G$, i.e. charge and temperature will not be discussed. The most clear meaning of these fundamental constants has the speed of light which relates two spacial dimensions of our daily experience: the time needed for traveling and the distance of travel. We do not travel with speed of light, of course, but $c$ is the accepted upper limit of traveling speed and independent of the observer. We can specify any speed as a fraction of the speed of light, or measure any distance by the time light would need to travel through. Planck's constant $\hbar$ is more difficult to interpret. It relates the energy of light to its frequency, or, if speed of light is included, to its wave length. Thereby, as speed of light relates length- and time-scale, Planck's constant relates energy- and time-scale. We accept Planck's constant $\hbar$  as the minimum `quantum of action' according to Heisenberg uncertainty principle. For now let us formulate the question: If $c$ and $\hbar$ only relate different scales of our physical world: What is the scale in itself? In the above mentioned canon of universal physical constants, the coefficient of gravitation $G$ fixes the scale of length, time or mass: Planck length $l_P = \sqrt{\frac {\hbar G}{ c^3}}$, Planck time $\tau_P = \sqrt{\frac {\hbar G}{c^5}}$ and Planck mass $M_P = \sqrt{\frac{\hbar c}G}$. It is argued, that our present understanding of physics breaks down below these scales, but they are definitely inaccessible by any means of experimentation in an anthropic environment. Who, however, has ever measured the gravitational constant $G$ outside of the solar system? Or even outside of earth? Measurements are difficult and subject to large scatter and systematic errors \cite{Rosi2014}. Therefore, assuming $G$ as a fundamental constant must be considered as a `reasonable hypothesis', rather than a fact. It is evident also that, if $G$ is not a fundamental constant, the `scale' of the universe by itself is a priori undetermined.   Although such a statement seems to knock over the foundations of physics, one may argue that it would be even more natural to accept that `no fundamental scale' exists, instead of that such a scale is `God-given'. If, on the other hand, no fundamental scale is given, how can we explain the empirical fact of typical scales of structures in the universe? 

\begin{figure}[ht]
 \centering
\includegraphics[width=8.5cm]{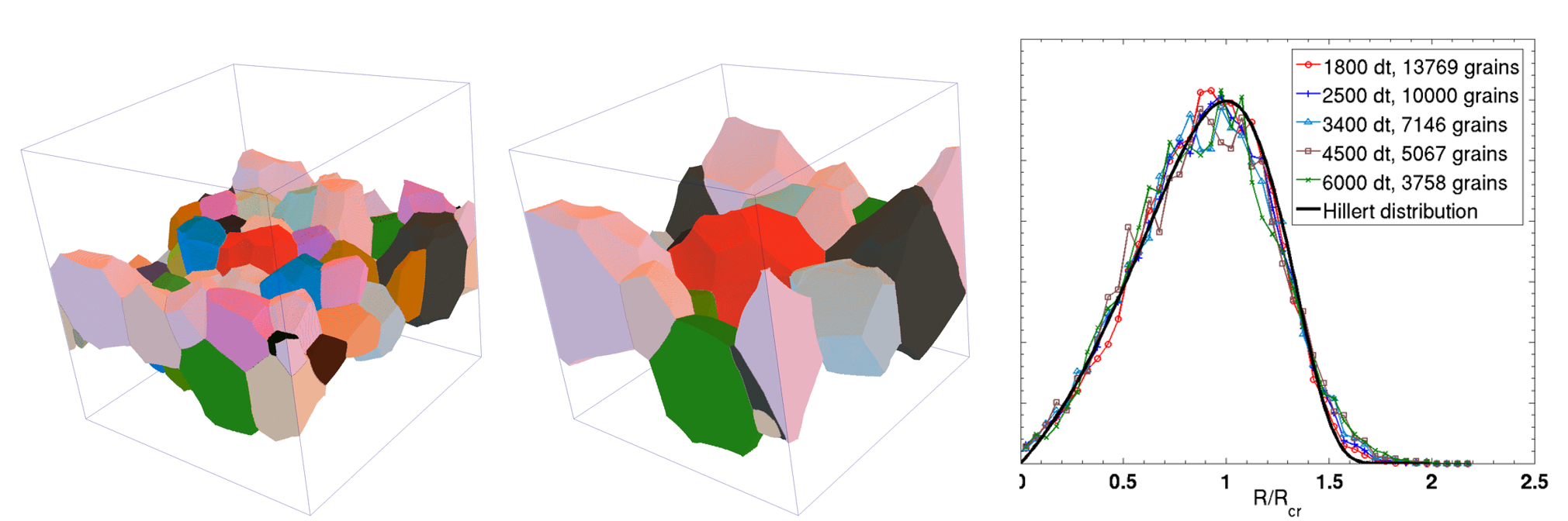}
\caption{Phase-field simulation of scale formation in a granular structure with increasing mean grain size (left), attaining a self similar distribution (right). From \cite{darvishikamachali2015}}\label{coarsening}
\end{figure}

Here we can learn from condensed matter physics, from the discipline of `pattern formation in mesoscopic systems' \cite{Zapolsky2015}. A `mesoscopic' system is defined small against a body where it is embedded in, that it feels no influence from macroscopic boundary conditions. The typical pattern arises from intrinsic mechanisms of self-organization. The scale of the system itself does not have to be stationary. It can be constantly evolving or approach some oscillating or chaotic limit cycle, as described by the logistic function \cite{Zehnder1994}. The only necessary condition for establishing a `scale' is, that characteristic features of the system are approaching some self-similar distribution in reduced coordinates. `Self-similar in reduced coordinates' denotes, that the size distribution function of objects $I$ of a size (or another characteristic scale) $\Omega_I$, evaluated relative to an appropriate mean size $\bar \Omega = \text{mean} (\{\Omega_I\})$, i.e. $\dfrac {\Omega_I}{\bar\Omega}$ becomes stationary. Examples are coarsening of foams or grains \cite{darvishikamachali2015} (see figure \ref{coarsening}) . The scale emerges from internal interactions in combination with appropriate conservation constraints. It shall be remarked that self-similarity of a size distribution and expansion of the characteristic scale are not in contradiction!

Recently, these concepts of mesoscopic systems have been applied by the authors to `pattern formation in the universe', where gravitational interaction is shown to be emergent from the action of quantum fluctuations in finite space objects (quantum-phase-fields)\cite{Steinbach2017zn,Steinbach2020,Kundin2020_zna}. A short summary of the concept will be presented in section \ref{QPF}. An important outcome of the analysis is a modified law of gravitation which predicts repulsive action on ultra-long distances, beyond a marginal distance $\Omega_\text{Earth} \approx 100 Mpc$, comparable to the size of the voids determined by observatory astronomy (see e.g. \cite {Mueller2000}). %Here it shall be remarked again that we do not consider the coefficient of gravitation as a fundamental constant, but emerging from laws of self organization (see section \ref{} for details). 

Our estimation of the marginal distance above is based on observations from earth, and the measured gravitational constant on earth (which itself is not treated as a constant but subject to the distribution of masses seen from the local observer). While in the limiting case of short distances (compared to $\Omega_\text{Earth}$) Newton's law of gravitation is recovered, repulsive action for distances larger than $\Omega_{\textit earth}$ rationalizes accelerating expansion \cite{Riess1998}.  Such a prediction is, of course, consistent with Einstein's general relativity concept with a small, positive cosmological constant $\Lambda = \frac {8\pi G}{c^2} \rho_{\rm vac}$ with the vacuum energy density $\rho_{\rm vac}$ \cite{Einstein1916,Einstein1917,Riess1998}. While it is still unclear how such a constant can be justified and how to adjust this constant \cite{Hebecker2000}, the appearance of repulsive gravitational action in the present concept is a mere consequent of energy conservation local form. We will proceed as follows:

\begin{itemize}
\item Recall basic features of quantum-phase-fields.
\item Derive vacuum energy from a special wave solution in the framework of the de~Broglie-Bohm (dBB) double solution program.
\item Construct a closed `doublon network' as the basic structure of the universe.
\item Discuss consequences for scale formation in an expanding universe.

\end{itemize}

\section{Ontology of space and mass based on quantum-phase-fields}\label{QPF}

A `quantum-phase-field' is an object which represents the basic categories of physical being and their relations. It has a substantial character in the sense, that it cannot be generated or destroyed. It is commonly termed `energy' in the sense of the first law of thermodynamics (1st Law) as the conserved category \cite{Lieb1999}. All manifestations of being thereby have to originate from an antisymmetric process of decomposition from the homogeneous state of `nothing' \footnote{Here we anticipate that `nothing' is the only amount of a substance which needs no creation.}. The stage of decomposition introduces a second category of being: `time' with a  given direction, related to entropy production, the second law of thermodynamics (2nd Law) \cite{Lieb1999}. Form this principle we take the statement that a symmetry broken state cannot be reverted to the original symmetric state, i.e. the symmetry breaking introduces a topological asymmetry, but does not violate 1st Law. 

The first stage of decomposition is the separation of positive states of energy $U_i$ and negative states of energy $E_I$, $i=1...n$, $I=1...N$ which have to sum to $0 = \sum_{i=1}^n U_i + \sum_{I=1}^N E_I$.  According to 2nd Law we will attribute the elements $U_i$ and $E_I$ with different characteristics in a common network where the respective elements separate / connect the respective other elements. Without proof we postulate the following topology as the simplest form consistent with the above principles:

\begin{itemize}
\item One $E$-element $E_K \Doteq E_{km}$ connects two U-elements $U_k$ and $U_m$, $k, m \in 1...n$.
\item One $U$-element $U_k$ connects a number of $M > 2$, $M \le N$ $E$-elements $E_K$, $K \in 1...M$.
\end{itemize}

\begin{figure}[ht]
 \centering
\includegraphics[width=5.0cm]{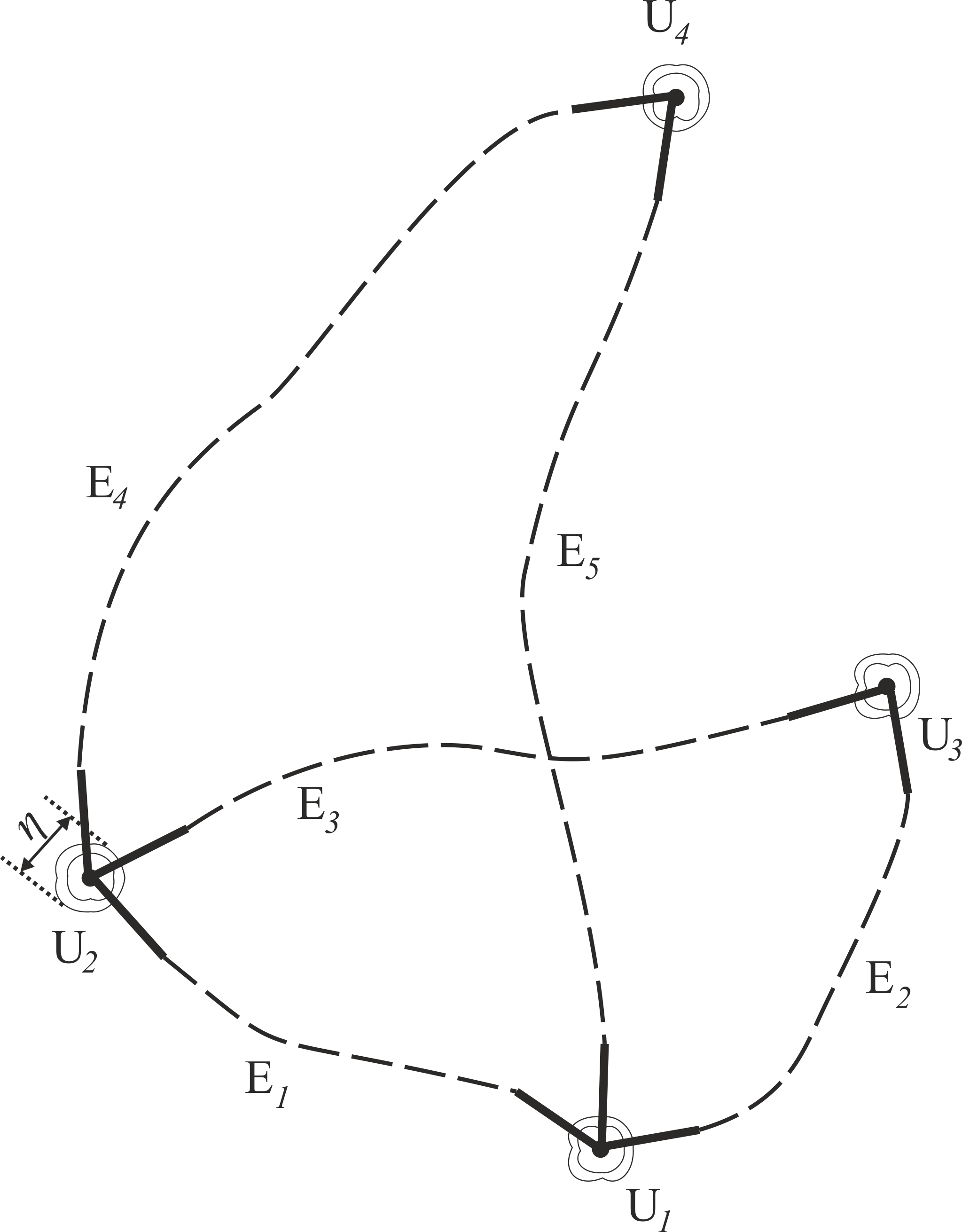}
\caption{Network of 4 $U$-elements (particles) with size $\eta$ and 5 $E$-elements (dashed lines, edges). This 3-dimensional space filling configuration may relate to a neutron-neutrino, or proton-electron pair. It is `closed in itself', i.e. there is no `loose end'. Though it needs nor necessary to be `complete', i.e. there may be missing edges , like the missing edge between node $U_3$ and $U_4$. In this case theses nodes are `invisible' to each other, though they are detectable via the edges with common nodes.}\label{network}
\end{figure}

Obviously this leads to a network as depicted in figure \ref{network} where the $U$-elements are `nodes' and the $E$-elements are `edges', for details see \cite{Kundin2020_zna}. 
Without loss of generality we associate the $U$-elements with positive energy, called `masses' or `particles', and the $E$-elements with negative energy, called `space'. Additional attributes, like charge, may be associated to further symmetry relations, but are not included in the present state of development of the concept. 

It is noteworthy that the connectivity of the network does not necessarily have to be complete. In other words, not all particles ($U$-elements) have to be connected to each other in general. An important property of the model also is that the energy associated with the connection of two particles always depends on a one-dimensional object, the $E$-element connecting two particles, and is thus independent of the dimensionality of the space-time in which the objects are embedded.

%\subsection{Quantum-phase-field}

Thus, we start from a set of elements $U_i$ and $E_I$, which fall into different categories and are ordered by a network topology. Both elements can formally be described by one mathematical object, called `quantum-phase-field' $\phi_I=\phi_I(s)$ defined on a 1-dimensional space coordinate $s$. This space coordinate is related to the E-element $E_{km}$ connecting two U-elements $U_k$ and $U_m$ according to the above detailed topological relations. The quantum-phase-field $\phi_I(s)$ represents both $U$- and $E$-states. $U$-states will be related to gradients of the field with positive energy within the Ginsburg-Landau (GL) formalism \cite{Landau1959}, where the Hamiltonian is expanded as function of the fields $\phi_I$. Also we will include time $t$, $\phi_I(s) \rightarrow  \phi_I(s,t)$. Time $t$ here is a sequence of configurations of the set of states, and we will not discuss the quest of time related to transportation of information. The set of fields $\phi_I(s,t)$ forms an isomorphism to the set of energetic states $U_i(t)$ and $E_I(t)$. They are the basic ingredients  to draw the physical world in a wave-mechanical picture, as outlined in section \ref{section_wave_guidance}. We will use these fields in two different respects. Firstly in section \ref{section_wave_guidance} we treat the fields as waves in the de~Broglie wave mechanical pictures and particles as realistic objects in Bohms interpretation. Second we invert the picture to a doublon network description in section \ref{doublon_network}. In this picture the waves are confined between the particles which form edges between 1-dimensional fields. Quantization follows canonically from the finiteness of spaces related to the fields $\phi_I$ \cite{Olsen1974}.

\section{Particle motion by wave guidance}\label{section_wave_guidance}

Here we relate to the isomorphism between the algebra of the set of $U$- and $E$-elements of energy and their topological relations, defining space and time, and the set of fields $\phi_I(s,t)$. The fields allow us to derive principal relations about the universe related to our anthropic perception of space and time. They relate to acknowledged concepts in physics, like the de~Broglie-Bohm (dBB) double solution program \cite{deBroglie1971}. In this framework we consider the structure of the Universe from the view in which particles are related to $U$-elements embedded in $E$-elements. The latter are the quantum statistical waves which obey the Klein-Gordon or Schr\"odinger equations. The energy of the $E$-elements can be defined to be positive or negative, but energy has no absolute scale in this case.

In our previous work \cite{Kundin2020_zna}, we have defined a superwave, $\Phi=a_\Phi e^{i\psi}$, as a solution of a phase-filed equation derived from a GL Hamiltonian, where $a_\Phi$ is an amplitude and $\psi$ is a phase. We have shown that there is the analogy to the particle motion in the water wave (the $U$-element embedded in the $E$-element). The velocity of the super wave is $v_\Phi = -\dfrac{\partial_t \Phi}{\partial_s \Phi}$, where $s$ is the direction of the wave propagation. By using equations $\partial_t \Phi = \dfrac{\partial_t a_\Phi}{a_\Phi}\Phi-\dfrac{v_\psi}{\eta}\Phi$, $\partial_s \Phi = \dfrac{i\Phi}{\eta}-\dfrac{1}{\eta}\psi\Phi\approx\dfrac{i\Phi}{\eta}$, for $\psi \ll 1$, and  $\dfrac{\partial_s a_\Phi}{a_\Phi}=\dfrac{1}{\eta}$, where $\eta$ is the interface width, we obtain $v_\Phi = v_{a_\Phi}+i v_\psi\psi$, where  $v_{a_\Phi}= \dfrac{\partial_t a_\Phi}{\partial_s a_\Phi}$. In the paper \cite{Kundin2020_zna} it was shown that $v_{a_\Phi}=\dfrac{c^2}{v_\psi}$ is the traveling velocity of the particle guided by the wave $\psi$. Here we show that this model has a direct analogy to the particle motion in the water waves, see also discussion in \cite{Willox2018}.

%Then the results which follows from this analogy will help us to explain the expansion of the universe due to the cosmological constant.

Let us assume a virtual direction $z$, which is normal to the particle traveling direction $s$ (alternatively one may consider longitudinal density variations like in acoustic wave propagation). The particle moves due to the traveling wave. By analogy to water waves we derive a velocity potential $ \psi_v(s,z,t)$ where $s$ is the line constant of the field and and $\nabla = \frac \partial {\partial s} \vec e_s + \frac \partial {\partial z} \vec e_z$ with orthogonal unit vectors $\vec e_s$ and $\vec e_z$. 
\begin{align}
\mathbf v_{\rm p} = \nabla \psi_v.
\end{align}
Then the continuity equation for the constant density model is
\begin{align}
\nabla\cdot \mathbf v_{\rm p} = \nabla^2 \psi_v=0 \text{  at  } z<\zeta,
\end{align}
where $\zeta$ is the surface profile function. The components of the particle velocity are defined as
\begin{align}
v_{\rm p,s} =\frac{\partial s}{dt }=\frac{\partial \psi_v}{ds }; \,\, v_{\rm p,z} =\frac{\partial \zeta}{dt }=\frac{\partial \psi_v}{dz }.
\end{align}

According to the simplified Bernoulli's equation, we can write 
\begin{align} \label{Bern}
\frac{\partial \psi_v}{\partial t} = -g\zeta,
\end{align}
where $g$ is the acceleration due to a virtual force in $z$-direction by analogy to the gravity.

By differentiating eq.  \eqref{Bern} we obtain
\begin{align}
\frac{\partial^2 \psi_v}{\partial t^2} = -g\frac{\partial \zeta}{\partial t} =-g v_{\rm p,z}=-g\frac{\partial \psi_v}{\partial z}.
\end{align}
The full set of equation for $\phi_v$ reads
\begin{align}
\frac{\partial^2 \psi_v}{\partial s^2} &+\frac{\partial^2 \psi_v}{\partial z^2}=0 & \text{  at  }  z<0,\\
\frac{\partial^2 \psi_v}{\partial t^2} &+g\frac{\partial \psi_v}{\partial z}=0 & \text{  at }  z=0, \label{Phi_equation_b}\\
\frac{\partial \psi_v}{\partial z} &= 0 & \text{  at  }  z\rightarrow-\infty.
\end{align}
A solution of equations is
\begin{align}
\psi_v(s,z,t) =Ae^{kz}\cos(ks-\omega t),
\end{align}
and the corresponding surface profile is
\begin{align}\label{Zeta_profile}
\zeta(s,z,t) =\eta\sin(ks-\omega t),
\end{align}
where $\eta=\dfrac{A\omega}{g}$ is a  length to be determined, $\omega$ is the angular frequency, $k=\dfrac{2\pi}{\lambda}$ is the wave number, $\lambda$ is the wave length. From eq. \eqref{Phi_equation_b}, $\omega = \sqrt{gk}$ because $-\omega^2 \cos(ks-\omega t) + g k \cos(ks-\omega t) = 0$, and from eq. \eqref{Zeta_profile} the wave velocity and the wave length are related by $v_\zeta = -\dfrac{\partial_t \zeta}{\partial_s \zeta}=\dfrac{\omega}{k} $. 
From the velocity potential the coordinates of the particle velocity are defined as follows:
\begin{align}
v_{\rm {p,} s}&=-\eta\omega e^{kz}\sin (ks-\omega t)\\
v_{\rm {p,} z}&=\eta\omega e^{kz}\cos (ks-\omega t).
\end{align}
Using a Taylor series about the point $s_0=0,z_0=0$ up to second order, the following traveling velocity in $s$ direction can be obtained
\begin{align}\label{PartVel}
v_{\rm {p,} s}=-\eta\omega \sin (ks_0+\omega t) + \frac{\eta^2\omega^2}{v_\zeta}.
\end{align}
Here the second term on the right hand side provides the drift velocity of the particle in the direction of the wave propagation.

 From the dBB theory \cite{deBroglie1971}, the traveling velocity of the particle due to the guidance by a wave $\psi$  is defined as $v_{\rm p}^{\rm trav}=\dfrac{c^2}{v_\psi}$, where $c$ is the the speed of light and $v_\psi= -\dfrac{\partial_t \psi}{\partial_s \psi}$ is the wave velocity. Since the second term on the right-hand side of eq. \eqref{PartVel} is the traveling velocity of the particle due to the guidance by the wave $\zeta$, we can deduce that $\eta\omega=c$.
Finally, the particle velocity is defined as
\begin{align}
v_{\rm {p,} s}=-c \sin (ks_0- \omega t) + \frac{c^2}{v_\zeta},
\end{align}
where the first term is the oscillation around point $s_0$ and the second term is the traveling velocity $v_{\rm p}^{\rm trav}$.

Since the $\zeta$ wave can be considered as the guidance wave in the dBB theory, it should obey the  Klein-Gordon equation  in relativistic case:
\begin{align}\label{KGeq}
\frac{1}{c^2}\frac{\partial^2\zeta}{\partial t^2}- \frac{\partial^2 \zeta}{\partial s^2} =\frac{m^2c^2}{\hbar^2} \zeta,
\end{align}
where $m$ is the mass of a particle. The Hamiltonian from which this equation is derived is similar to the Hamiltonian of a phase-field equation for a relativistic singularity \cite{Kundin2020_zna}, where $\zeta$ corresponds to a phase-field variable $\psi$, which is a guiding wave. More precisely, the phase-field equation has the form of the Klein-Gordon equation with advection. The corresponding  potential for the phase-field variable is equal to $f(\psi)= \frac{m^2c^2}{2\hbar^2}\psi^2$.

According to eq. \eqref{KGeq}, the angular frequency and the wave number of $\zeta$ are related by
\begin{align}
\frac{\omega^2}{c^2} = k^2 +\frac{m^2c^2}{\hbar^2}.
\end{align}
From this, we can define the acceleration by
\begin{align}
g = \dfrac{\omega^2}{k} = c^2k +\dfrac{m^2c^4}{\hbar^2k}.
\end{align}
Using this relations we can find the limits of the traveling velocity:
\begin{itemize}
 \item if $v_{\rm p}^{\rm trav}\rightarrow0$, then $v_\zeta\rightarrow\infty$, $k\rightarrow 0$ (because the momentum is 0), $\lambda\rightarrow \infty$, $\omega\rightarrow \dfrac{mc^2}{\hbar}$,  $g\rightarrow \infty$, $\eta\rightarrow \dfrac{\hbar}{mc}$.
  \item if $v_{\rm p}^{\rm trav}\rightarrow c$, then $v_\zeta\rightarrow c$, $k\rightarrow \infty$, $\lambda\rightarrow 0$, $\omega\rightarrow \infty$, $g\rightarrow \infty $, $\eta\rightarrow 0$.
\end{itemize}

 In the non-relativistic case one should use the Schr\"odinger equation
    \begin{align}       
    \frac{\partial\zeta}{\partial t}=\frac{\hbar}{2mi} \frac{\partial^2 \zeta}{\partial s^2} +\frac{iU}{\hbar} \zeta. \,\,\left[=\frac{iE}{\hbar}\zeta\right]      
    \end{align}
    Here we define $E$ as the eigenvalue of the Hamilton operator in consistency with quantum mechanics. As shown in \cite{Kundin2020_zna} the Schr\"odinger equation can be obtained from the phase-field equation for a pilot wave with $U=U_0\left(1- \frac{c^2}{2v_\psi}\right)$ with $U_0=mc^2$.
    
    For a potential $U$, the standard dispersion relation reads
        \begin{align}       
        k=\frac{\sqrt{2m(E-U)}}{\hbar}.       
        \end{align}
   Hence
   \begin{align} \label{eq_g}
   g = \dfrac{\omega^2}{k}= \dfrac{E^2}{\hbar\sqrt{2m(E-U)}}.
\end{align}

In the limit $v_{\rm p}^{\rm trav}\rightarrow0$ we have $k\rightarrow 0$ because $v_{\rm p}^{\rm trav}\sim k$. In order to get $v_\zeta=\dfrac{\omega}{k}\rightarrow\infty$, we should have $\omega\rightarrow constant$. This can be achieved if $E\rightarrow U\rightarrow U_0$ in eq. \eqref{eq_g}. Hence, if $v_{\rm p}^{\rm trav}\rightarrow0$, then $v_\zeta\rightarrow\infty$, $k\rightarrow 0$ , $\lambda\rightarrow \infty$, $\omega\rightarrow \dfrac{U_0}{\hbar}=\dfrac{mc^2}{\hbar}$,  $g\rightarrow \infty$, $\eta=\dfrac{c}{\omega}\rightarrow \dfrac{c\hbar}{U_0}=\dfrac{\hbar}{mc}$.

  Here by the comparison to the relativistic Klein-Gordon equation, it can be seen that for the case $v_{\rm p}^{\rm trav}\rightarrow0$, $\eta\rightarrow \dfrac{\hbar}{mc}$.  Hence our assumption $|U_0|=mc^2$ is  valid. Note that the potential can be negative or positive according to eq.~\eqref{eq_g}.
    Finally, we can see that there exists a ground  state energy $U_0$ of a pilot wave corresponding to massive energy. The cause of this energy is given by the oscillations of the guiding waves even in the case of zero particle traveling velocity,inducing oscillations of particles around their stationary positions. To estimate the minimal amplitude of the oscillations, we introduce a minimal distance $\eta_{\rm min}=\dfrac{\hbar}{mc}$, at which a particle position can be defined. Note that in this relation, in comparison to the classical definition of the de~Broglie wave length $\lambda_{dB}=\dfrac{\hbar}{mv_p^{\rm trav}}$, the particle occupies a finite space even for $v_p^{\rm trav}\rightarrow 0$. 
    
\section{Closed Doublon Network}\label{doublon_network}

In the previous section a single field $\psi$ had been analyzed in regard to its individual components as the particle component $\zeta$. As said before, this analysis in the dBB wave mechanical framework can only be solved for systems with given boundary conditions, as in a quantum mechanical experiment like the double slit experiment where the experimental set up sets the boundary conditions. For discussion see \cite{Kundin2020_zna,Willox2018}. 

\begin{figure}[ht]
 \centering
\includegraphics[width=8.0cm]{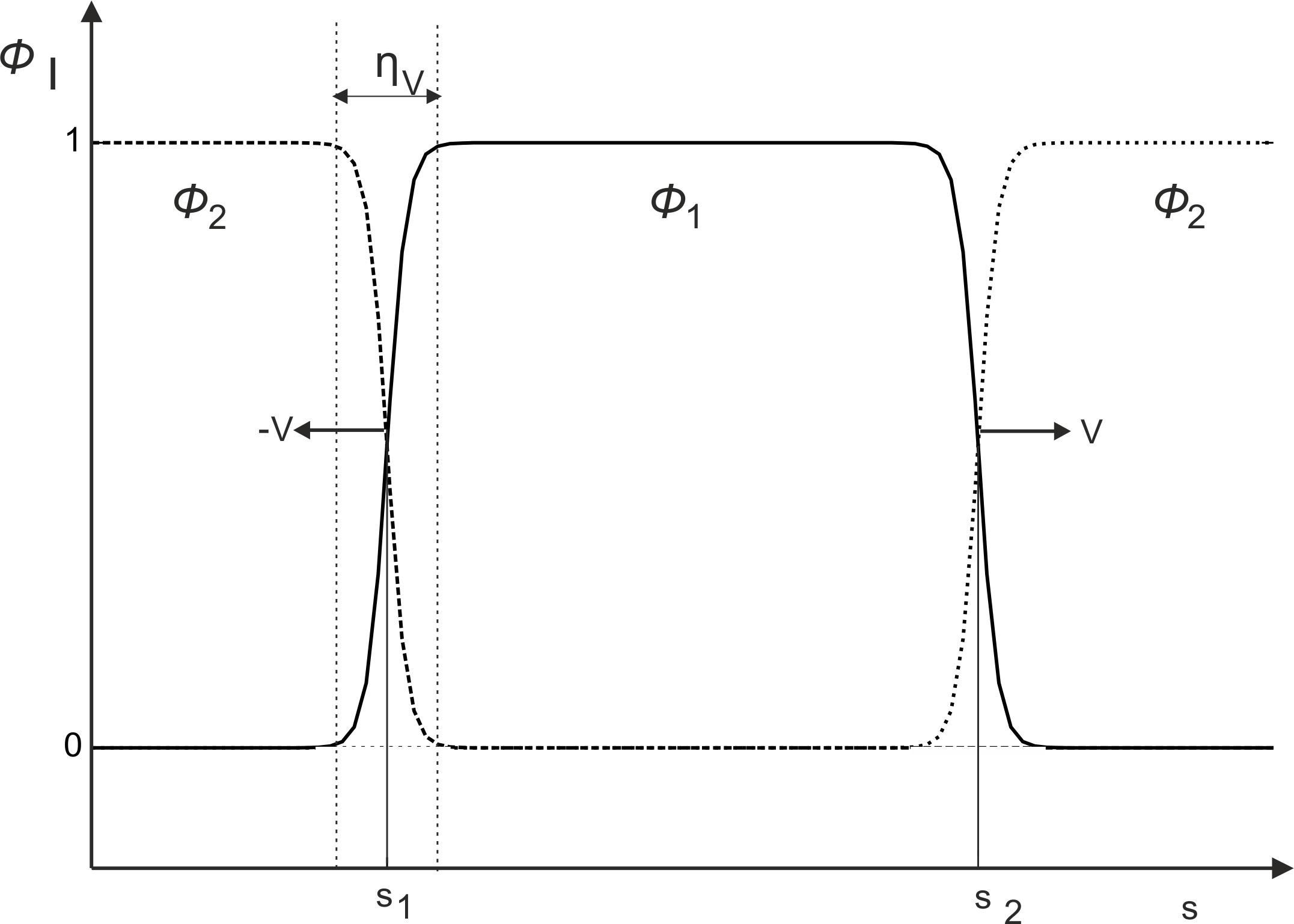}
\caption{Sketch of a pair of two doublons in a periodic setting, each of the doublons consisting of an anti-symmetric pair of half sided solitons \cite{Kundin2020_zna}.}\label{doublon}
\end{figure}

The identical formalism, however, can be applied in a `moving boundary' setting where the particles define the (moving) boundary conditions for the waves $\psi$ which follow a linear, Schr\"odinger type equation on the one hand. The particles are described by the solution of a phase-field, or soliton type, non-linear wave equation with a particle solution $\phi$. They are guided, on the other hand, by the $\psi$ waves they are connected to. The coupled system evolves according to internal mechanisms of self organization. The solutions of the wave equations are called `doublons' as a combination of two antisymmetric solitons \cite{Kundin2020_zna}. This solution is schematically depicted in figure \ref{doublon} for two single fields $\phi_1$ and $\phi_2$.  Each field is normalized by $0 \le \phi_I \le 1$. The state $\phi_I \equiv 0$ of an individual field has no physical meaning, $\phi_I \equiv 1$ represents the $E$-element while intermediate states $0 < \phi_I < 1$ correspond to $U$-elements, see figure \ref{network}. Different fields are connected by a conservation constraint $\sum_{I=1}^N  \phi_I =1$ as postulated in \cite{Steinbach1999}. Now we can reason this constraint from the second topological principle, that each $U$-element connects a number of $E$-elements, i.e. there are no `loose ends' of fields but all fields end in an $U$-element. By the sum constraint the set of fields is closed in itself, forming a `universe'. For details see \cite{Steinbach2017zn,Steinbach2020,Kundin2020_zna}.

The solution is depicted in figure \ref{doublon} for two particle waves $\phi_1$ and $\phi_2$, also called `doublon' since it consists of two half sided solitons.

The doublon forms a container of finite size for quantum oscillations $\psi$ with a discrete spectrum. 
The energy of this wave spectrum defines the apparent scale of the doublon $\Omega_{ij} = \dfrac {U_0 \eta} {E_{ij}}$ normalized by the energy quantum $U_0$ corresponding to the rest mass and the length quantum $\eta$. In the general case the density of vacuum fluctuations on the doublon, however, are not in equilibrium (equally distributed). Then the scales seen from different particles will not be equal $\Omega_{ij} \ne \Omega_{ji}$, where $\Omega_{ij}$ denotes the scale seen from particle $i$ and $\Omega_{ji}$ the scale seen from particle $j$. The energy $E_{ij}$ associated with the 'space' connecting the observer $i$ with a $U$-element $j$ is defined from the Casimir forces on a 1-dimensional edge between two nodes \cite{Casimir1948,Steinbach2017zn}:
\begin{align}\label{SpaceEnergyij}
E_{ij} =-\alpha_i \frac {m_j}{\Omega_{ij}}.
\end{align}
It is noteworthy that $E_{ij}\neq E_{ji}$ and depends on the observer $i$ via the coefficient $\alpha_i$ to be determined later. As will be shown below, $\alpha_i$ (and consequently the energy $E_{ij}$) depends on the set of space-like elements connecting $i$ to other particles. Since this set is in general different among observers, $\alpha_i$ depends on the specific observer. Without loss of generality, $m_j$ can be viewed as the mass of the particle $j$.

The energy $E_i =\sum_{j=1}^{N_i}E_{ij}$ attributed to the node $i$ connected to a number of nodes $j$ via the edges $ij$ is thus given by, 
\begin{align}\label{ParticleEnergy}
 E_i =-\alpha_i \sum _{j=1}^{N_i} \frac {m_j}{\Omega_{ij}}.
\end{align}
where $N_i$ is the number of fields, which meet at the particle or $U$-element $i$.
Here it must be noted that $E_i$ is a local quantity and no `instantaneous action at a distance' is necessary to evaluate the energy balance according to the principle of neutrality \cite{Steinbach2017zn,Steinbach2020} $E_i + p U_i =0$, where $p=1$ for periodic systems, i.e. where each two particles are connected twice, and $p=2$ for non-periodic systems \footnote{$p$ in this context needs not to be strictly an integer, since not all particles have to be connected, but may be used as an adjustable parameter to account for `dark matter'.}. We calculate the constant $\alpha_i$ using $E_i=-pU_i$ as
\begin{align}\label{AlphaClosed}
\alpha_i=p U_i \left[\sum _{j=1}^{N_i}\frac{m_j}{\Omega_{ij}} \right]^{-1}.
\end{align}

Further, we define a characteristic size, $\bar\Omega_i$, with respect to the locus $i$ as the geometric mean 
\begin{align}\label{harmonic}
\bar\Omega_i=N_i\left[\sum _{j=1}^{N_i}\frac{m_j}{\Omega_{ij} } \right]^{-1} \frac{\sum_j^{N_i} m_j}{N_i}.
\end{align}

Importantly, the scale $\bar \Omega_i$ is not a universal one but depends on the specific observer $i$ and the set of $E$-elements, which connect it to the 'world'. In the context of this essay we will only deal with quantities observed in the local environment of the earth at present time.

Using~\ref{harmonic}, we can  rewrite \eqref{AlphaClosed} as $\alpha_i=2U_i\bar\Omega_i/\sum_j m_j$ which after insertion in to (\ref{SpaceEnergyij}) gives
\begin{align}\label{ParticleEnergyClosed2}
E_{ij} =  - \frac{pU_i}{\tilde{N}_{ij} }\frac{\bar\Omega_i}{\Omega_{ij}},
\end{align}
where we define $\tilde N_{ij} =(\sum_k^{N_i} m_k)/m_j$.
This interaction energy depends on  the scale of the doublon $\Omega_{ij}$ and on the characteristic scale $\bar \Omega_i$. Both scales have the dimension of length, however they must not be considered as (relative) space coordinates in classical mechanics, since they depend on the position of the local observer (see above, for discussion see  \cite{Steinbach2017zn,Steinbach2020}). Only if the time $\tau_{ij}$ for equilibration of the vacuum fluctuation defining the scale $\Omega_{ij}$ is negligible compared to the inverse minimum frequency of fluctuation $\dfrac 1{\nu_{\textit min}}=\dfrac{\Omega_{ij}}c$, one can use the formal relation between the scales \eqref{harmonic} to determine the microscopic force acting on $j$ by the observer $i$, $f_{ij}^{\textit micro}$:

\bea\label{microforce}
\nonumber f_{ij}^{\textit micro} &=&-\left(\frac{\partial  E_{ij}}{\partial\Omega_{ij}}+ \frac{\partial  E_{ij}}{\partial\bar\Omega_{i}}\frac {\partial\bar\Omega_{i}}{\partial\Omega_{ij}} \right ) \\
&=& \frac{pU_i\bar\Omega_{i}}{\tilde N_{ij}}\frac {1}{\Omega_{ij}^2}\left[1-\frac{\bar\Omega_{i}}{\tilde N_{ij} \Omega_{ij}}\right].
\eea

For distances $\Omega_{ij} < \tilde N_{ij} \bar\Omega_{i}$ we note repulsive action while for $\Omega_{ij} \gg \tilde N_{ij} \bar\Omega_{i}$ Newton's law of gravity is recovered. Identifying the prefactor in (\ref{microforce}) with $G m_i m_j$ ($G$=the coefficient of gravitation), one obtains
\begin{equation}\label{G-via-Omegabar}
\bar\Omega_i=\dfrac{Gm_im_j\tilde N_{ij}}{pU_i}=\dfrac{GM_i}{pc^2},
\end{equation}
where we used $U_i=m_ic^2$ and defined the mass of the universe, seen by the 'observer' $U_i$, to be $M_i=m_j\tilde N_{ij}=\sum_k^{N_i}m_k$. Note that the mass of the universe depends on the specific observer, $i$. When measured from a basis on Earth, the gravitational constant is given by $G~\approx~6.67~e^{-11} \left [\frac {m^3}{kg s^2} \right ]$. Further, for the mass of universe, measured by an observed on Earth, we set $M_\text{Earth}\approx~10^{52}$kg  \cite{Persinger2009}. We thus obtain an estimate of the scale of the universe viewed from the Earth $\bar\Omega_{i} = \bar\Omega_\text{Earth}$:
\be\label{size}
\bar\Omega_\text{Earth} \approx 7.4 \times10^{24}\, {\rm m} \approx  \frac 1p 240\, {\rm Mpc},
\ee
which is, for a non periodic system, $p=2$ consistent with observations of the size of the large voids in the universe. Regarding the meaning of the mass, $m_j$, we set for the number of particles in the observable universe $N\approx 10^{79}$ and obtain $m_j=M/N\approx 10^{-27}$ kg. This is roughly equal to the mass of a proton.

Instantaneous `action at a distance' and repulsive gravitational action at the micro-scale is thereby limited to distances $\frac {\bar\Omega_{\textit earth}}N$ below the Plack lenth. We leave closer consideration to future work.

On `cosmological' distances, one must clearly treat $\bar\Omega_{i}$ and $\Omega_{ij}$ as independent, since the cosmological scale $\bar\Omega_{i}$ cannot be directly related to a length defined in a current microscopic environment, or even the solar system. Variation of the energy of space \eqref {ParticleEnergyClosed2} with respect to $\bar\Omega_{i}$ and $\Omega_{ij}$ independently gives:

\bea\label{ForceClosed}
\nonumber f_{ij} &=&-\left(\frac{\partial  E_{ij}}{\partial\Omega_{ij}}+ \frac{\partial  E_{ij}}{\partial\bar\Omega_{i}}\right) \\
&=&\frac{pU_i\bar\Omega_{i}}{\tilde N_{ij}\Omega_{ij}^2}\left[1-\frac{\Omega_{ij}}{\bar\Omega_{i}}\right].
\eea

For illustration we treat the limit of 3 particles with $\tilde N_{ij}=2$. There are three possible cases:
\begin{itemize}
 \item if $\Omega_{ij}=\Omega_{ik}$, $\bar\Omega_{i}=\Omega_{ij}$, then  $f_{ij}  =0$,
 \item if $\Omega_{ij}\ll\Omega_{ik}$, $\bar\Omega_{i}=2\Omega_{ij}$, then  $f_{ij} = \dfrac{pU_i}{2\Omega_{ij}}$ 
 
 and $f_{ik} = -\dfrac{pU_i}{2\Omega_{ik}}$,
  \item if $\Omega_{ij}\gg\Omega_{ik}$, $\bar\Omega_{i}=2\Omega_{ik}$, then $ f_{ij} = -\dfrac{pU_i}{2\Omega_{ij}}$ 
  
  and $f_{ik} = \dfrac{pU_i}{2\Omega_{ik}}$.
\end{itemize}

In the last case, if the distance between the particles is much larger then the average value dominated by two close particles due to the homogeneous mean, the forces becomes repulsive.

Treating the system as periodic, $p=1$, we can correlate this length to the size of large voids \cite{Mueller2000} which seems reasonable at least in the visible universe. Also a non-periodic system $p=2$ lies in the span of the large uncertainties of these estimations. Therefore it is hard to draw a conclusion on periodicity from these analogies. Structures beyond the marginal distance $\bar\Omega_\text{Earth}$ repel each other, leading to an accelerating expansion. In the limit $\Omega_{ij} \gg \bar\Omega_\text{Earth}$ we further see from the generalized gravitational law \eqref{ForceClosed} that the force scales as $f_{ij} \propto - \dfrac 1 {\Omega_{ij}}$ instead of $f_{ij} \propto  \dfrac 1 {(\Omega_{ij})^2}$ for small distances, i.e. that repulsive gravitational action on ultra-long distances decays more slowly with distance than attractive gravitational action on small distances. %We further see that, because $\bar\Omega_{i}$ is defined as the harmonic mean of the distances $\Omega_{ij}$ \eqref{harmonic}, it's value is dominated by small distances $\Omega_{ij} \ll \bar \Omega_{i}$. This means that $\Omega_{i}$ approaches a constant for the limit of ultra large distances $\Omega_{ij} \gg \bar\Omega_\text{Earth}$ and the universe may be considered isotropic on these scales. Furthermore, we see that the force now scales $f_{ij} \propto \frac 1 {\Omega_{ij}}$, i.e. that repulsive gravitational action on ultra-long distances decays more slowly with distance than attractive gravitational action on small distances i.e. the related spacial energy has logarithmic divergence in the limit $\Omega_{ij} \rightarrow \infty$. 
Consequences of this statement deserve further considerations in the future.

\section{Discussion and conclusion}\label{DC}

In section \ref{section_wave_guidance} we considered a model of the universe in a wave mechanical picture according to the dBB double solution program with a guiding wave, oscillating in a virtual direction $z$. These waves are related to particle waves, which oscillate with the same phase as the guiding wave and move in analogy to particles on water waves. If the traveling velocity of a particle is not zero, the corresponding wave has a finite phase velocity moving in the direction of the particle on the line coordinate $s$. The oscillations of waves have the minimum energy $U_0$ corresponding to the massive energy of the particles. Assigning this energy to the pilot wave instead of the particle wave allows us to motivate a kind of uniformly distributed `the ground state potential' with a total energy density in the visual universe $NU_0/V_H$, where $V_H$ is the  Hubble volume and $N$ is the number of  particles. 

According to the standard cosmological model of expanding universe, the energy conservation equation for a particle of mass $m$ moving from a center of mass $M$ with the velocity $v$ can be written as
\begin{align}\label{EnergyConservation}
\frac{m v^2}{2}- \frac{mMG}{a}- Nmc^2\frac{4\pi a^2\eta}{V_H}=0,
\end{align}
where $a$ is the distance between the particle and the center,  also known as scale factor, $4 \pi a^2\eta$ is a volume of a thin layer which is responsible for the ground state potential of one particle.
Then by substituting $v=\dot a$,  $\dfrac {4\pi}3 \rho_{\rm U} =\dfrac M {a^3}$, $\eta =\dfrac{\bar\Omega_{\textit earth}}{N}$ \cite{Steinbach2017zn}, and $V_H=\dfrac{4 \pi R_H^3}{3}$, where $R_H$ is the Hubble radius, we obtain 
\begin{align}\label{EnergyConservation2}
\frac{\dot a^2}{a^2}- \frac{8\pi G}{3 } \rho_{\rm U}-\frac{3  c^2 \Omega_{\textit earth}}{R_H^{3}}=0.
\end{align}
It can be seen that if the ground state potential is negative it should cause the acceleration of the expansion of the universe.
The comparison with the cosmological constant gives $\Lambda\cong9\dfrac{ \Omega_{\textit earth}}{R_H^{3}}\cong 1.2\times  10^{-52}$ m$^{-2}$. This value is close to the value defined   as $\Lambda =3\dfrac{H_0^2}{c^2}\Omega_\Lambda=3\dfrac{\Omega_\Lambda}{R_H^2}$,  where  $H_0$ the measured Hubble constant and $\Omega_\Lambda$ is the ratio between the energy density due to the cosmological constant and the critical density. Hence the last term in eq.~\eqref{EnergyConservation2} gives the  acceleration behavior similar to the cosmological constant for todays values of $\Omega_{\textit earth}$ and  $R_H$.

On the other hand,  $U_0$ is related to the  mass of a particle, hence we can estimate the total energy density related to $U_0$ of all particles in universe using the mass density of visual universe $\rho_{\rm U} = 3\times 10^{-28}$ kg/m$^3$ 

This is a factor $20$ smaller than the vacuum energy density $\rho_{\rm vacuum} = 5.96\times 10^{-27}$ kg/m$^3$ defined by the standard model of cosmology, using $\Lambda = \dfrac{8\pi G}{c^2}\rho_{\rm vacuum}$ \cite{Planck_2016}. Hence, we can state that the ground state potential can be the rational of a cosmological constant $\Lambda$ in the case or an open universe as described by the dBB double solution program. The factor $20$ between the massive energy and the estimated `vacuum energy' of the universe is well in agreement with common assumptions about dark energy. The new point here is the interpretation in the wave mechanical picture of the dBB double solution program, where we have the traditional understanding of massive particles embedded in an `open space'. We may argue that this picture overestimates energetic contributions related to space largely compared to energy related to mass. This has to be corrected by introducing `dark energy' in order to be consistent with observations.

Reverting the picture to a closed doublon network model, section \ref{doublon_network}, gives reasonable agreement without introducing an additional ground state energy. Here the prediction of repulsive gravitational action on ultra-long distances is a consequence of neutrality and energy conservation, which leads to the repulsive term in the gravitational force between two bodies \eqref{ForceClosed}. Nevertheless, we can show in this picture an analogy to the standard model or cosmological expansion, if we define from \eqref{ForceClosed} the coefficient of gravitation dependent on the distances $\tilde G(\Omega_{ij},\bar \Omega_\text{Earth}) = G_0 \left[ 1 - \dfrac {\Omega_{ij}}{\bar \Omega_\text{Earth}}\right]$. The acceleration becomes:
\be\label{acceleration}
 \frac{\ddot a}a=-\frac {M \tilde G}{a^3}=-\frac {4\pi}3 \rho_{\rm U} G_0\left[1- \frac {a} {\bar \Omega_\text{Earth}}\right].
\ee
In the case of self similarity the last term in \eqref{acceleration} should approach a constant, i.e.  $\bar \Omega_\text{Earth} \propto a$. The model predicts an accelerating expansion $\dfrac{\ddot a}a \propto \rho_{\rm U}$ which should fade out proportionally to $\dfrac1{a^3}$ for constant mass of the universe.  We leave closer consideration to future work.

In conclusion, we have outlined a wave mechanical framework of the `universe' along the double solution approach by de~Broglie-Bohm with wave representation of particles and space. The traditional dBB picture was extended to a model of the particles `swimming' on the pilot wave by analogy to the particles in the dense water waves. The particles have a traveling velocity related to quantum oscillations of the pilot wave. This model predicts the existence of a ground state energy of the quantum oscillations with a finite vacuum energy rationalizing the assumption of a cosmological constant in the standard model of cosmology. Reverting this picture we define a closed doublon network , where particles form the boundary for vacuum fluctuation along the 1-dimensional path between particles. According to Casimir \cite{Casimir1948} we define the negative vacuum energy of the quantum fluctuations. Finally, we have shown that the gravitational force \eqref{ForceClosed}, resulting from this model can be displayed in a form consistent with a positive cosmological constant.  Finally we state that a self similarity of the distribution of masses in the universe, corresponding to a constant relation between scale factors $a\propto \bar\Omega_\text{Earth}$, is not in contradiction to the observation of an ever expanding universe.

%\section*{Acknowledgement} 

\newpage
%\section{References}
\bibliography{./Universe_Scale}

\begin{thebibliography}{10}

\bibitem{Rosi2014}
G.~Rosi F.~Sorrentino L.~Cacciapuoti~M. Prevedelli and G.M. Tino.
\newblock Precision measurement of the newtonian gravitational constant using
  cold atoms.
\newblock {\em Nature}, 510:518--21, 2014.

\bibitem{Zapolsky2015}
H.~Zapolsky.
\newblock {\em Kinetics of Pattern Formation: Mesoscopic and Atomistic
  Modelling}, pages 153--192.
\newblock 05 2015.

\bibitem{Zehnder1994}
H.~Hofer and E.~Zehnder.
\newblock {\em Symplectic Invariants and Hamiltonian Dynamics}.
\newblock 1994.

\bibitem{darvishikamachali2015}
R.~D. Kamachali, A.~Abbondandolo, K.~F. Sieburg, and I.~Steinbach.
\newblock Geometrical grounds of mean field solutions for normal grain growth.
\newblock {\em Acta Materialia}, 90, 2015.

\bibitem{Steinbach2017zn}
I.~Steinbach.
\newblock Quantum-phase-field concept of matter: Emergent gravity in the
  dynamic universe.
\newblock {\em Z. Naturforschung A}, 72:51–58, 2017.

\bibitem{Steinbach2020}
I.~Steinbach.
\newblock Erratum to: Quantum-phase-field concept of matter: Emergent gravity
  in the dynamic universe.
\newblock {\em Zeitschrift fur Naturforschung A}, pages 89--91, 2020.

\bibitem{Kundin2020_zna}
J.~Kundin and I.~Steinbach.
\newblock Quantum-phase-field: from de broglie--bohm double solution program to
  doublon networks.
\newblock {\em Z. Naturforschung A}, 75(2):155--170, 2020.

\bibitem{Mueller2000}
V.~M\"uller, S.~Arbabi-Bidgoli, J.~Einasto, and D.~Tucker.
\newblock Voids in the las campanas redshift survey versus cold dark matter
  models.
\newblock {\em Mon. Not. R. Astron. Soc.}, 318:280--288, 2000.

\bibitem{Riess1998}
A.G.~Riess et~al.
\newblock Observational evidence from supernovae for an accelerating universe
  and a cosmological constant.
\newblock {\em Astronomical J.}, 116:1009--38, 1998.

\bibitem{Einstein1916}
A.~Einstein.
\newblock Die {G}rundlage der allgemeinen {R}elativit\"atstheorie.
\newblock {\em Annalen der Physik}, 49:769--822, 1916.

\bibitem{Einstein1917}
A.~Einstein.
\newblock Kosmologische {B}etrachtungen zur {A}llgemeinen
  {R}elativit\"tstheorie.
\newblock {\em Sitzungsberichte der Königlich Preußischen Akademie der
  Wissenschaften (Berlin)}, page 142–152, 1917.

\bibitem{Hebecker2000}
A.~Hebecker and C.~Wetterich.
\newblock Quintessential adjustment of the cosmological constant.
\newblock {\em Phys. Rev. Lett}, 85:3339--3342, 2000.

\bibitem{Lieb1999}
E.H. Lieb and J.~Yngvason.
\newblock The physics and mathematics of the second law of thermodynamics.
\newblock {\em Physics Reports}, 310:1--96, 1999.

\bibitem{Landau1959}
L.D. Landau and E.M. Lifshitz.
\newblock {\em Statistical Physics Part 1, third revised edition 1980}.
\newblock Pergamon, Oxford, 1959.

\bibitem{Olsen1974}
S.O. Olsen.
\newblock A natural way of quantization.
\newblock {\em Acta Physica Academiae Scientiarum Hungaricae}, 37:97--103,
  1974.

\bibitem{deBroglie1971}
L.~de~Broglie.
\newblock L’interpretation de la mechanique ondulatoire par la theorie de la
  double solution.
\newblock {\em Proc. Znt. Sch. Phys. Enrico Fermi}, 49:346--367, 1971.

\bibitem{Willox2018}
S.~Colin M.~Hatifi, R.~Willox and T.~Durt.
\newblock Bouncing oil droplets, de broglie's quantum thermostat, and
  convergence to equilibrium.
\newblock {\em Entropy}, 20:32, 2018.

\bibitem{Steinbach1999}
I.~Steinbach and F.~Pezzolla.
\newblock A generalized field method for multiphase transformations using
  interface fields.
\newblock {\em Physica~D}, 134:385--393, 1999.

\bibitem{Casimir1948}
H.~Casimir.
\newblock On the attraction between two perfectly conducting plates.
\newblock {\em Proc. Kon. Nederland. Akad. Wetensch.}, B51:793--795, 1948.

\bibitem{Persinger2009}
M.A. Persinger.
\newblock A simple estimate for the mass of the universe: Dimensionless
  parameter {A} and the construct of "pressure".
\newblock {\em J. of Physics Astrophysics and Physical Cosmology}, 3:1--3,
  2009.

\bibitem{Planck_2016}
Planck collaboration (2016).
\newblock Planck 2015 results. xiii. cosmological parameters.
\newblock {\em Astronomy \& Astrophysics}, page 594:A13, 2016.

\end{thebibliography}
%\begin{thebibliography}{10}
%\end{thebibliography}
\bibliographystyle{unsrt}

\end{document}